\documentclass[a4paper,11pt]{article}
\usepackage[utf8]{inputenc}
\usepackage[english]{babel}
\usepackage{amsmath, amssymb,graphicx}
\pdfoutput=1
\usepackage{jheppub}

\title{Simple rules of functional integration in the Schwarzian theory: SYK correlators }

\author[a,b]{Vladimir V. Belokurov}
\author[a]{and Evgeniy T. Shavgulidze}

\affiliation[a]{Lomonosov Moscow State University,\\ Leninskie gory 1, Moscow, 119991, Russia}
\affiliation[b]{Institute for Nuclear
Research of the Russian Academy of Sciences,\\ 60th October Anniversary
Prospect 7a, Moscow, 117312, Russia}

\emailAdd{vvbelokurov@yandex.ru}
\emailAdd{shavgulidze@bk.ru}

\abstract
{ We derive the general rules of functional integration in the theories of the Schwarzian type, and evaluate explicitly the functional integrals assigning correlation functions in the SYK model.}


\arxivnumber{1811.11863v2}

\begin{document}
\maketitle
\section{Introduction}
\label{sec:intr}

The appearance of a dynamics common to different physical models in different fields may indicate the existence of a hidden symmetry behind it.
It has recently become clear that a quantum mechanical model of Majorana fermions with a random interaction (Sachdev-Ye-Kitaev model), the holographic
description of the Jackiw-Teitelboim dilaton gravity, open string theory and some other models lead to the same effective  theory with the Schwarzian action
\begin{equation}
   \label{Act2}
  I_{Schw}= -\frac{1}{\sigma^{2}}\int \limits _{S^{1}}\,\left[ \mathcal{S}_{\varphi}(t)+2\pi^{2}\left(\varphi'(t)\right)^{2}\right]dt\,,
\end{equation}
where $S^{1} $ is the unit circle, and
\begin{equation}
   \label{Der}
\mathcal{S}_{\varphi}(t)=
\left(\frac{\varphi''(t)}{\varphi'(t)}\right)'
-\frac{1}{2}\left(\frac{\varphi''(t)}{\varphi'(t)}\right)^2
\end{equation}
is the Schwarzian derivative.

The list of the papers includes (but is not confined to) \cite{(SY)} - \cite{(Sachdev)}, and it will not rest here.

In some approximations, these physical models appear to be reparametrization invariant. And $SL(2,\textbf{R})$-invariant action (\ref{Act2}) inherits the emergent reparametrization symmetry.

Let us remind briefly how the Schwarzian action originates in the SYK model and in 2D JT gravity.
Functional integrals in the SYK model can be rewritten in terms of the bilocal bosonic functions $\tilde{G} $ and $\tilde{\Sigma} $ with the effective bosonic action
$A_{eff}(\tilde{G},\,\tilde{\Sigma})$
$$
\int\,F(\tilde{G},\,\tilde{\Sigma})\,\exp\{-A_{eff}(\tilde{G},\,\tilde{\Sigma}) \}\,d\tilde{G}\,d\tilde{\Sigma}\,.
$$
The saddle points of the infrared essential part of the effective bosonic action $A_{eff}$
\begin{equation}
   \label{Saddle}
  G_{f}(t_{1},\,t_{2})=\frac{\left(f'(t_{1})f'(t_{2})\right)^{\frac{1}{4}}}{\left | f(t_{2})-f(t_{1})\right|^{\frac{1}{2}}}
\end{equation}
form the space $\mathcal{F}$ of one-time differentiable functions $f$ on the unit circle.

The rest (infrared inessential) part of the effective bosonic action $A_{eff}$ results (for three-time differentiable $f$) in the factor
$$
\exp\left\{\frac{1}{\sigma^{2}}\int \limits _{S^{1}}\, \left[\left(\frac{f''(t)}{f'(t)}\right)'
-\frac{1}{2}\left(\frac{f''(t)}{f'(t)}\right)^2\right]\,dt\,  \right\}
$$
in the integrand.

Thus the averaged saddle point solution (or two-point SYK correlation function) has the form
\begin{equation}
   \label{Gav}
 < G_{f}(t_{1},\,t_{2})>=\int \limits_{\mathcal{F}}\,\frac{\left(f'(t_{1})f'(t_{2})\right)^{\frac{1}{4}}}{\left |f( t_{2})-f(t_{1})\right|^{\frac{1}{2}}}
 \exp\left\{\frac{1}{\sigma^{2}}\int \limits _{S^{1}}\, \mathcal{S}_{f}(t)\,dt\, \right\}\,df\,.
\end{equation}

The action of Jackiw-Teitelboim gravity is
\begin{equation}
   \label{JT}
A_{JT}=Topological\ term-\alpha\left(\int d^{2}x\sqrt{g}\,\phi\,(R+2)+2\int \limits_{bdy}\sqrt{\gamma} \,\phi\, K \right)\,,
\end{equation}
where $\phi$ is the dilaton field and $K$ is the extrinsic curvature.
While the equation of motion for dilaton demands the space to be $AdS_{2}\,,$ variations of the closed boundary curve with the proper boundary condition for $\phi$ transform
the last term in the action into
$$
\frac{1}{\sigma^{2}}\int \limits _{S^{1}}\, \mathcal{S}_{f}(t)\,dt\,,
$$
where $t$ is a parameter of the closed boundary curve.

Functional integrals in these theories are the integrals  with the measure
\begin{equation}
   \label{MeasureS}
   \mu_{\sigma}(d\varphi)=\exp\left\{\frac{1}{\sigma^{2}}\int \limits _{S^{1}}\, \mathcal{S}_{\varphi}(t)\,dt  \right\}  d\varphi
\end{equation}
over the group of diffeomorphisms $ Diff^{1}_{+}(S^{1})$ .

In \cite{(BShExact)}, \cite{(BShCorrel)}, we explicitly evaluated some nontrivial functional integrals of this type using the quasi-invariance of the measure $\mu\,.$

Although one can use the invariance of the Schwarzian  theory to link it to another theory where the corresponding calculations are much simpler than in the original one \cite{(BAK1)} - \cite{(SW)}, \cite{(BShUnusual)}, \cite{(BShPolar)}, we consider it essential to elaborate a special technique of functional integration in the Schwarzian theory.

Here, we would like to emphasize that there is a significant difference between functional integration over the group $ Diff^{1}_{+}(S^{1})$ and that over the group $Diff^{1}_{+}(\textbf{R})\,.$

The group $SL(2, \textbf{R})$ is turned into the subgroup of $ Diff^{1}_{+}(S^{1})$ by linear-fractional transformation
$$
\exp \{ i2\pi\psi\}=\frac{\alpha\exp\{i2\pi\phi \}-\beta}{\bar{\beta}\exp\{i2\pi\phi \}-\bar{\alpha}}\,,\ \ \ \ \ |\alpha|^{2} - |\beta|^{2}=1\,.
$$

The measure (\ref{MeasureS}) is $SL(2, \textbf{R})$ invariant.

Therefore to obtain finite results in case of $SL(2, \textbf{R})$ invariant integrands, we should
integrate over the quotient space $Diff^{1}_{+}(S^{1})/SL(2,\textbf{R})\,. $

However $SL(2, \textbf{R})$ is not a subgroup of the group  $Diff^{1}_{+}(\textbf{R})\,.$
For example, the transformation $ g_{a}\circ f=g_{a} (f) = (f+a)^{-1}\,\  \left(g_{a}\in SL(2, \textbf{R})\,,\ f\in Diff^{1}_{+}(\textbf{R})\right)$ removes $ g_{a}\circ f $ from $Diff^{1}_{+}(\textbf{R})\,.$

At the same time, the subgroup $ P=A\,N $ of $SL(2, \textbf{R})$ consisting of transformations $f\rightarrow af+b$ (see \cite{(Lang)} Ch. III, par. 1) is a
subgroup of  $Diff^{1}_{+}(\textbf{R})\,.$
The group $P$ is noncompact, and for invariant integrands, one should factorize the integration space, that is one should consider the functional integrals over the quotient space  $Diff^{1}_{+}(\textbf{R})/ P\,.$

Thus it is quite natural that the functional integrals over different integration spaces give different results for the same integrand and determine different quantum theories.

In this paper, we complete the elaboration of the functional integrals calculus with the measure (\ref{MeasureS}).

In section \ref{sec:quasi-inv}, we review and refine the approach to functional integration based on the quasi-invariance of the measure, and define functional integration over the quotient space $Diff^{1}_{+}(S^{1})/SL(2,\textbf{R})\,. $

In section
\ref{sec:div}, we derive the simple universal rules of functional integration in the theories of the Schwarzian type and demonstrate how the functional integrals are reduced to ordinary multiple integrals.
Although the method is developed for functional integrals over the space $Diff^{1}_{+}(S^{1})/SL(2,\textbf{R})\,, $ it can also be modified to evaluate functional integrals
over the space $Diff^{1}_{+}(\textbf{R})/P\,. $

In section \ref{sec:two} and in appendices B, C,
we evaluate the functional integrals assigning correlation functions in the corresponding models.

   In section \ref{sec:concl}, we discuss the difference between the results in these two cases and give the concluding remarks.

\section{Quasi-invariance of the measure as a key to functional integration}
\label{sec:quasi-inv}

Invariant measures analogous to the Haar measure on finite-dimensional groups do  not exist for the noncompact groups $H$ \cite{(Weil)}. However, there can exist measures that are quasi-invariant with respect to the action of a more smooth subgroup $G\subset H.$ The quasi-invariance means that under the action of the subgroup $G$ the measure transforms to itself multiplied by a function $  \mathcal{R}_{g}(h)$ (the Radon-Nikodim derivative) parametrized by the elements of the subgroup $g\in G\,. $

The measure (\ref{MeasureS}) on the group of diffeomorphisms $ Diff^{1}_{+}(S^{1})$ is quasi-invariant with respect to the action of the subgroup  $Diff^{3}_{+}(S^{1})$ \cite{(Shavgulidze1978)}-\cite{(Shavgulidze2000)}.

 To evaluate functional integrals over the group $Diff^{1}_{+} (S^{1})\,,$ it is convenient to represent them as integrals over the group $Diff^{1}_{+}([0,1])$ with the ends of the interval $[0,\,1]$ glued $(\varphi'(0)=\varphi'(1))\,.$
After fixing a point $t=0$ on the circle of unit length $S^{1}$, the integral $\int \limits _{S^{1}}$ is written as $\int \limits _{0}^{1}\,.$

The measure on the group of diffeomorphisms of the interval $Diff^{1}_{+}([0,1])\  ( \varphi(0)=0\,,\, \varphi(1)=1\,,\  \varphi'(t)>0 )$ has the form \cite{(Shavgulidze1978)}-\cite{(Shavgulidze2000)}
\begin{equation}
   \label{Measure}
   \mu_{\sigma}(d\varphi)=\frac{1}{\sqrt{\varphi'(0)\varphi'(1)}} \exp\left\{ \frac{1}{\sigma^{2}}\left[   \frac{\varphi''(0)}{\varphi'(0)}-  \frac{\varphi''(1)}{\varphi'(1)}\right]\right\}\exp\left\{\frac{1}{\sigma^{2}}\int \limits _{0}^{1}\, \mathcal{S}_{\varphi}(t)\,dt  \right\}  d\varphi\,.
\end{equation}

Now, the integral  over $Diff^{1}_{+} (S^{1})$ turns into the the integral over $Diff^{1}_{+}([0,1])$ as follows \cite{(BShCorrel)}:
\begin{equation}
   \label{Eq:Equality}
\frac{1}{\sqrt{2\pi}\sigma}\int\limits_{Diff^{1}_{+} (S^{1}) }F(\varphi)\mu_{\sigma}(d\varphi)
=\int\limits_{Diff^{1}_{+} ([0,1]) }\delta\left(\frac{\varphi'(1)}{\varphi'(0)}-1 \right)\,F(\varphi)\,\mu_{\sigma}(d\varphi)
\,.
\end{equation}

The measure $\mu_{\sigma}(d\varphi) $ is generated by the Wiener measure under some special substitution of variables \cite{(Shavgulidze1978)}-\cite{(Shavgulidze2000)} (see also \cite{(BSh)}).
Namely, if we consider  a continuous function on the interval $[0,\,1]\ \xi(t)$ satisfying the boundary condition
$\xi(0)=0\ \left(\xi\in C_{0}([0,\,1])\, \right)\,,$
then under the substitution
\begin{equation}
   \label{subst}
 \varphi(t)=\frac{\int \limits _{0}^{t}\,e^{\xi(\tau)}d\tau}{\int \limits _{0}^{1}\,e^{\xi(\eta)}d\eta }  \,,\ \ \ \ \ \xi(t)=\log\varphi'(t)-\log\varphi'(0)\,,
\end{equation}
the measure $\mu_{\sigma}(d\varphi)$ on the group $ Diff^{1}_{+}([0,\,1])$ turns into the Wiener measure  $w_{\sigma}(d\xi)$ on $C_{0}([0,\, 1])\,.$
In this way, we get the following equality of functional integrals
\begin{equation}
   \label{IntEq}
  \int\limits_{Diff^{1}_{+} ([0,\,1]) }F(\varphi)\,\mu_{\sigma}(d\varphi)
 =\int\limits _{C_{0}([0,1]) }F(\varphi(\xi))\, w_{\sigma}(d\xi)\,.
\end{equation}

After the substitution $\ f(t)= - \cot \pi\varphi (t)$
in (\ref{Gav}), $< G_{f}(t_{1},\,t_{2})> $ is rewritten in the form
$$
 < G(t_{1},\,t_{2})>=\sqrt{2\pi}\sigma\,\int \limits_{Diff^{1}_{+}([0,\,1])}\delta\left(\frac{\varphi'(1)}{\varphi'(0)}-1 \right)
$$
\begin{equation}
   \label{GavFI}
\times\,\frac{\left(\varphi'(t_{1})\varphi'(t_{2})\right)^{\frac{1}{4}}}{\left |\sin \left[\pi \varphi ( t_{2})-\pi\varphi(t_{1})\right]\right|^{\frac{1}{2}}}
 \exp\left\{\frac{ 2\pi^{2}}{\sigma^{2}}\int \limits _{0}^{1}\,\left(\varphi'(t)\right)^{2}\,dt\, \right\}\,\mu_{\sigma}(d\varphi)\,.
\end{equation}

The quasi-invariance of the measure  (\ref{Measure}) with respect to the action of the subgroup  $Diff^{3}_{+}([0,\,1])$ gives
  $$
  \int \limits _{Diff^{1}_{+}([0, 1])}\,F(\varphi)\mu_{\sigma}(d\varphi)=\frac{1}{\sqrt{g'(0)g'(1)}}\int \limits _{Diff^{1}_{+}([0, 1])}\,F(g(\varphi))
  $$
  \begin{equation}
   \label{FI1}
  \times \exp\left\{ \frac{1}{\sigma^{2}}\left[   \frac{g''(0)}{g'(0)}\varphi'(0)-  \frac{g''(1)}{g'(1)}\varphi'(1)\right]   +        \frac{1}{\sigma^{2}}\int \limits _{0}^{1}\, \mathcal{S}_{g}\left(\varphi(t)\right)\,\left(\varphi' (t)\right)^{2}\,dt  \right\} \,\mu_{\sigma}(d\varphi)\,.
\end{equation}

In what follows, we assume the function $g$ to be
\begin{equation}
   \label{g}
g(t)=g_{\alpha}(t)=\frac{1}{2}\left[ \frac{1}{\tan\frac{\alpha}{2}}\tan\left(\alpha(t-\frac{1}{2}) \right)+1 \right]\,.
\end{equation}
In this case,
\begin{equation}
   \label{tan}
g'_{\alpha}(0)=g'_{\alpha}(1)=\frac{\alpha}{\sin \alpha}\,,
\ \ \ -\frac{g''_{\alpha}(0)}{g'_{\alpha}(0)}=\frac{g''_{\alpha}(1)}{g'_{\alpha}(1)}=2\alpha\tan\frac{\alpha}{2}\,, \ \ \ \mathcal{S}_{g_{\alpha}}(t)=2\alpha^{2}\,,
\end{equation}
and the equation (\ref{FI1}) looks like:
$$
\int \limits _{Diff^{1}([0, 1])}F(g_{\alpha}(\varphi))\,\exp\left\{-\frac{2\alpha}{\sigma^{2}}\tan\frac{\alpha}{2}\left(\varphi'(0)+\varphi'(1)\right)\right\}
 \exp\left\{\frac{2\alpha^{2}}{\sigma^{2}}\int \limits _{0}^{1}\left(\varphi' (t)\right)^{2}dt  \right\} \mu_{\sigma} (d\varphi)
$$
\begin{equation}
   \label{FI2}
= \frac{\alpha}{\sin \alpha}\,\int \limits _{Diff^{1}([0, 1])}F(\varphi)\,\mu_{\sigma}(d\varphi)\,.
\end{equation}

Denote
$$
\Psi (\varphi)=\exp\left\{-\frac{2\alpha}{\sigma^{2}}\tan\frac{\alpha}{2}\left(\varphi'(0)+\varphi'(1)\right)\right\}\,F(g_{\alpha}(\varphi))\,.
$$
Then
\begin{equation}
   \label{FPsi}
F(\varphi)=\exp\left\{\frac{2\alpha}{\sigma^{2}}\tan\frac{\alpha}{2}\left(\left(g^{-1}_{\alpha}(\varphi)\right)'(0)
+\left(g^{-1}_{\alpha}(\varphi)\right)'(1)\right)\right\}\,\Psi \left(\left(g^{-1}_{\alpha}(\varphi)\right)\right)\,.
\end{equation}
Thus the integral
\begin{equation}
   \label{FI3A}
J^{\alpha}=\int \limits _{Diff^{1}([0, 1])}\,\Psi(\varphi)\,
 \exp\left\{\frac{2\alpha^{2}}{\sigma^{2}}\int \limits _{0}^{1}\left(\varphi' (t)\right)^{2}dt  \right\}\, \mu_{\sigma} (d\varphi)
\end{equation}
is transformed into
$$
J^{\alpha}=\frac{\alpha}{\sin \alpha}
$$
\begin{equation}
   \label{FI3}
\times\int \limits _{Diff^{1}([0, 1])}\exp\left\{\frac{2\alpha}{\sigma^{2}}\tan\frac{\alpha}{2}\left(\left(g^{-1}_{\alpha}(\varphi)\right)'(0)
+\left(g^{-1}_{\alpha}(\varphi)\right)'(1)\right)\right\}\Psi \left(\left(g^{-1}_{\alpha}(\varphi)\right)\right)\mu_{\sigma}(d\varphi)\,.
\end{equation}

For the function $g_{\alpha} $ given by (\ref{g}), the inverse function is
$$
\left(g_{\alpha}^{-1}(\varphi)\right)(t)=\frac{1}{\alpha}\arctan\left[\tan \frac{\alpha}{2}\,\left(2\varphi(t)-1\right)\right]+\frac{1}{2}\,,
$$
and
$$
\left(g^{-1}_{\alpha}(\varphi)\right)'(0)=\frac{\sin \alpha}{\alpha}\,\varphi'(0)\,,\ \ \ \ \ \left(g^{-1}_{\alpha}(\varphi)\right)'(1)=\frac{\sin \alpha}{\alpha}\,\varphi'(1)\,,
$$
$$
\exp\left\{\frac{2\alpha}{\sigma^{2}}\tan\frac{\alpha}{2}\left(\left(g^{-1}_{\alpha}(\varphi)\right)'(0)
+\left(g^{-1}_{\alpha}(\varphi)\right)'(1)\right)\right\}=\exp\left\{\frac{4\,\sin ^{2}\frac{\alpha}{2}}{\sigma^{2}}\left(\varphi'(0)+\varphi'(1)\right)\right\}\,.
$$

For the factor
$$
\frac{\left(\varphi'(t_{1})\varphi'(t_{2})\right)^{\frac{1}{4}}}{\left |\sin \left[\pi \varphi ( t_{2})-\pi\varphi(t_{1})\right]\right|^{\frac{1}{2}}}
$$
in the integrand of (\ref{GavFI}) , we have
\begin{equation}
   \label{F1}
\lim\limits_{\alpha\rightarrow \pi -0}\frac{\left(\left(g^{-1}_{\alpha}(\varphi)\right)'(t_{1})\left(g^{-1}_{\alpha}(\varphi)\right)'(t_{2})\right)^{\frac{1}{4}}}
{\left |\sin \left[\pi \left(g^{-1}_{\alpha}(\varphi)\right) ( t_{2})-\pi\left(g^{-1}_{\alpha}(\varphi)\right)(t_{1})\right]\right|^{\frac{1}{2}}}
=\pi^{-\frac{1}{2}}\,\frac{\left(\varphi'(t_{1})\varphi'(t_{2})\right)^{\frac{1}{4}}}{\left |\varphi ( t_{2})-\varphi(t_{1})\right|^{\frac{1}{2}}}
\end{equation}

Now the functional integral for the two-point SYK correlation function (\ref{GavFI}) is transformed into
$$
 < G_{2}(t_{1},\,t_{2})>_{SYK}^{\alpha}=\sqrt{2}\sigma\,\frac{\alpha}{\sin \alpha}
$$
 \begin{equation}
   \label{2GSYK}
 \times\int \limits_{Diff^{1}_{+}([0,\,1])} \delta\left(\frac{\varphi'(1)}{\varphi'(0)}-1 \right)\, \frac{\left(\varphi'(t_{1})\varphi'(t_{2})\right)^{\frac{1}{4}}}{\left |\varphi ( t_{2})-\varphi(t_{1})\right|^{\frac{1}{2}}}
 \exp\left\{\frac{ 4}{\sigma^{2}}\left(\varphi'(0)+\varphi'(1)\right) \right\}\,\mu_{\sigma}(d\varphi)+ O(1)\,.
\end{equation}

Note that the integral (\ref{2GSYK}) is invariant under the shifts of the variable $t$ (that is, under the choice of the point on the $S^{1}$ corresponding to the left end of the interval $[0,\,1]$):
$$
 < G_{2}(t_{1},\,t_{2})>_{SYK}^{\alpha}\,=\, < G_{2}(0,\,t_{2}-t_{1})>_{SYK}^{\alpha}\,.
$$

Generally speaking, the functional integrals (\ref{FI3})  converge for $ 0< \alpha < \pi \,, $ and diverge for $ \alpha = \pi \,. $
The integral (\ref{2GSYK}) is a typical example.
The point is that the Schwarzian action is invariant under the noncompact group of linear-fractional transformations.
To get the finite results for functional integrals in the Schwarzian theory, one should factor the infinite input of this noncompact group out.

Technically, it is convenient to exclude the input of the $SL(2, \textbf{R})$ group that is a subgroup of $Diff^{1}_{+}(S^{1})\,, $ or to
integrate over the quotient space $Diff^{1}(S^{1})/SL(2,\textbf{R})\,. $

To this end, we propose first to evaluate regularized $(\alpha<\pi)$ functional integrals over the group  $Diff^{1}([0, 1])$ (with the ends glued) and then normalize them to the corresponding integrals over the group $SL(2, \textbf{R})\,.$
Thus, we define the renormalized functional integral  $J^{R}$  as the limit
\begin{equation}
   \label{JR}
J^{R}=\lim \limits_{\alpha\rightarrow\pi - 0}\  \frac{ J^{\alpha}}{ \int \limits _{SL(2, \textbf{R})}\,\Psi(\varphi_{z})\,
 \exp\left\{\frac{2\alpha^{2}}{\sigma^{2}}\int \limits _{0}^{1}\left(\varphi'_{z} (t)\right)^{2}dt  \right\}\,d\nu_{H}}\,.
\end{equation}
Here, $ J^{\alpha}$ is given by (\ref{FI3A}), $\varphi_{z}\in SL(2, \textbf{R})$ and $d\nu_{H}$ is the invariant Haar measure on $SL(2, \textbf{R})\,.$

The factor
$$
\frac{\left(\varphi'(t_{1})\varphi'(t_{2})\right)^{\frac{1}{4}}}{\left | \varphi ( t_{2})-\varphi(t_{1})\right|^{\frac{1}{2}}}
$$
in the integrand of (\ref{2GSYK}) is $SL(2, \textbf{R})$ invariant.

Therefore, for renormalized  SYK correlation functions we have
\begin{equation}
   \label{GRen}
<G>^{R}_{SYK}=\lim \limits_{\alpha\rightarrow\pi - 0}\  \frac{ <G>^{\alpha}_{SYK}}{ V^{\alpha}_{SL(2, \textbf{R})}}\,,
\end{equation}
where $ V^{\alpha}_{SL(2, \textbf{R})}$ is the regularized volume of the group $SL(2,\textbf{R})$ \cite{(BShCorrel)}:
$$
V^{\alpha}_{SL(2, \textbf{R})}=\int \limits _{SL(2,\textbf{R})}\exp \left\{-\frac{2\left[ \pi^{2}-\alpha^{2}\right]}{\sigma^{2}}\,\int \limits _{0}^{1}(\varphi'(t))^{2}dt  \right\}  d\nu_{H}=\frac{\pi\sigma^{2}}{\pi^{2}-\alpha^{2}}\,\exp\left\{-\frac{2\left(\pi^{2} - \alpha^{2} \right)}{\sigma^{2}} \right\}\,.
$$

Thus, the renormalization excludes the singularity $ (\pi - \alpha)^{-1}$  in the functional integrals (\ref{FI3}), and we
get the finite results for SYK correlation functions.

\section{ Splitting the interval  multiplies functional integrals}
\label{sec:div}

In the case when the functional $F$ in  (\ref{IntEq}) has the special form
$$
F(\varphi)=\Phi(\varphi(t_{1});\,\varphi'(0),\,\varphi'(t_{1}),\,\varphi'(1))\,,
$$
the equation (\ref{IntEq}) looks like
$$
\int\limits_{Diff^{1}_{+} ([0,\,1]) }\,\Phi(\varphi(t_{1});\,\varphi'(0),\,\varphi'(t_{1}),\,\varphi'(1))\,\mu_{\sigma}(d\varphi)
$$
\begin{equation}
   \label{IntEqFi1}
=\int\limits _{C_{0}([0,1]) }\,\Phi(\varphi(\xi(t_{1}));\,\varphi'(0),\,\varphi'(\xi(t_{1})),\,\varphi'(\xi(1)))\, w_{\sigma}(d\xi)\,.
\end{equation}

It is convenient to split the interval $[0,\,1]$ into the two intervals $[0,\,t_{1}]$ and $[t_{1},\,1]\,,$ and substitute
\begin{equation}
   \label{xieta}
\xi(t)=\eta_{0}\left(\frac{t}{t_{1}} \right)\,,\ \ \ 0\leq t\leq t_{1}\,; \ \ \ \ \ \ \ \ \ \ \ \xi(t)=\eta_{0}(1) +\eta_{1}\left(\frac{t-t_{1}}{1-t_{1}} \right)\,,\ \ \ t_{1}\leq t \leq 1\,.
\end{equation}
Now the exponent of the Wiener measure is written as
$$
\frac{1}{\sigma^{2}}\left(\int \limits _{0}^{t_{1}}\left(\xi'(t) \right)^{2}\,dt +\int \limits _{t_{1}}^{1}\left(\xi'(t) \right)^{2}\,dt\right)=\frac{1}{\sigma^{2}t_{1}}\int \limits _{0}^{1}\left(\eta'_{0}(t) \right)^{2}\,dt+\frac{1}{\sigma^{2}(1-t_{1})}\int \limits _{0}^{1}\left(\eta'_{1}(t) \right)^{2}\,dt\,.
$$
Due to the Markov property of  the Wiener process $\xi(t) \,,$ the measure $w_{\sigma}(d\xi)$ turns into the product of the two measures
$$
w_{\sigma \sqrt{t_{1}}}(d\eta_{0})\,w_{\sigma \sqrt{1-t_{1}}}(d\eta_{1})\,.
$$

To return to the integrals over the group of diffeomorphisms, define the functions $ \psi_{0}\,,\ \psi_{1}\ \in Diff^{1}_{+}([0,\,1])$
\begin{equation}
   \label{psi}
 \psi_{0}(t)=\frac{\int \limits _{0}^{t}e^{\eta_{0}(\tau)}d\tau }{\int \limits _{0}^{1}e^{\eta_{0}(\tau)}d\tau }\,, \ \ \ \ \ \ \ \ \ \ \ \ \psi_{1}(t)=\frac{\int \limits _{0}^{t}e^{\eta_{1}(\tau)}d\tau }{\int \limits _{0}^{1}e^{\eta_{1}(\tau)}d\tau }\,.
\end{equation}

In this way, we get
$$
\int\limits_{Diff^{1}_{+} ([0,\,1]) }\,\Phi(\varphi(t_{1});\,\varphi'(0),\,\varphi'(t_{1}),\,\varphi'(1))\,\mu_{\sigma}(d\varphi)
$$
\begin{equation}
   \label{IntEqFi2}
=\int\limits_{Diff^{1}_{+} ([0,\,1]) }\,
\int\limits_{Diff^{1}_{+} ([0,\,1]) }\,\Phi(\varphi(t_{1});\,\varphi'(0),\,\varphi'(t_{1}),\,\varphi'(1))\,\mu_{\sigma\sqrt{t_{1}}}(d\psi_{0})\,\mu_{\sigma\sqrt{1-t_{1}}}(d\psi_{1})
\,.
\end{equation}

The functions $\varphi$ and $\psi$ (and their derivatives) are related by the following equations:
\begin{equation}
   \label{FiPsi0}
\varphi(t)=\frac{t_{1}\,\psi'_{1}(0)\,\psi_{0}\left(\frac{t}{t_{1}} \right)}{t_{1}\psi'_{1}(0)+(1-t_{1})\psi'_{0}(1)}\,,\ \ \ \ \ 0\leq t\leq t_{1}\,;
\end{equation}
\begin{equation}
   \label{FiPsi1}
\varphi(t)=\frac{t_{1}\,\psi'_{1}(0)+(1-t_{1})\,\psi'_{0}(1)\,\psi_{1}\left(\frac{t-t_{1}}{1-t_{1}} \right)}{t_{1}\psi'_{1}(0)+(1-t_{1})\psi'_{0}(1)}\,,\ \ \ \ \ t_{1}\leq t \leq 1\,,
\end{equation}
and
\begin{equation}
   \label{DerFiPsi0}
\varphi'(t)=\frac{\,\psi'_{1}(0)\,\psi'_{0}\left(\frac{t}{t_{1}} \right)}{t_{1}\psi'_{1}(0)+(1-t_{1})\psi'_{0}(1)}\,,\ \ \ \ \ 0\leq t\leq t_{1}\,;
\end{equation}
\begin{equation}
   \label{DerFiPsi1}
\varphi'(t)=\frac{\psi'_{0}(1)\,\psi'_{1}\left(\frac{t-t_{1}}{1-t_{1}} \right)}{t_{1}\psi'_{1}(0)+(1-t_{1})\psi'_{0}(1)}\,,\ \ \ \ \ t_{1}\leq t \leq 1\,.
\end{equation}

In particular, the arguments in the integrand are
\begin{equation}
   \label{FiT1}
\varphi(t_{1})=\frac{t_{1}\,\psi'_{1}(0)}{t_{1}\psi'_{1}(0)+(1-t_{1})\psi'_{0}(1)}\,;\ \ \ \ \ \ \
\varphi' (t_{1})=\frac{\psi'_{0}(1)\psi'_{1}(0)}{t_{1}\psi'_{1}(0)+(1-t_{1})\psi'_{0}(1)}\,,
\end{equation}
\begin{equation}
   \label{DerFi01}
\varphi' (0)=\frac{\psi'_{0}(0)\psi'_{1}(0)}{t_{1}\psi'_{1}(0)+(1-t_{1})\psi'_{0}(1)}\,,
\ \ \ \ \ \ \
\varphi' (1)=\frac{\psi'_{0}(1)\psi'_{1}(1)}{t_{1}\psi'_{1}(0)+(1-t_{1})\psi'_{0}(1)}\,.
\end{equation}

Note that the substitution (\ref{subst}) is nonlocal, and for diffeomorphisms $\varphi(t)$ the Markov property is not valid.

Now  we can represent all functional integrals with integrands depending on $\varphi(t_{1}),\ \varphi'(0),$ $ \varphi'(t_{1}),\ \varphi'(1)$  in a similar way. To this end, we define the basic functional integral:
\begin{equation}
   \label{E}
\mathcal {E}_{\sigma}(u,\,v)=\int\limits_{Diff^{1} ([0,1]) }\,\delta\left(\varphi'(0)-u \right)\,\delta\left(\varphi'(1)-v \right)\,\mu_{\sigma}(d\varphi)\,,
\end{equation}
and rewrite the functional integral (\ref{IntEqFi2}) as
$$
\int\limits_{Diff^{1}_{+} ([0,\,1]) }\,
\int\limits_{Diff^{1}_{+} ([0,\,1]) }\,\Phi\left(\varphi(t_{1});\,\varphi'(0),\,\varphi'(t_{1}),\,\varphi'(1)\right)\,\mu_{\sigma\sqrt{t_{1}}}(d\psi_{0})\,\mu_{\sigma\sqrt{1-t_{1}}}(d\psi_{1})
$$
\begin{equation}
   \label{IntUV01}
=\int\limits_{0}^{+\infty}\cdot\cdot\cdot
\int\limits_{0}^{+\infty}du_{0}dv_{0}du_{1}dv_{1} \Phi\left(\varphi_{uv}(t_{1});\,\varphi'_{uv}(0),\,\varphi'_{uv}(t_{1}),\,\varphi'_{uv}(1)\right)\,\mathcal {E}_{\sigma\sqrt{t_{1}}}(u_{0},\,v_{0})\,\mathcal {E}_{\sigma\sqrt{1-t_{1}}}(u_{1},\,v_{1})\,.
\end{equation}

Instead of the variables $(u_{0},\,v_{0},\,u_{1},\,v_{1})\,,$ it is convenient to use the variables
$$
z_{1}=\varphi(t_{1})\,,\ \ \ x_{0}= \varphi'(0)\,,\ \ \ y_{1}=\varphi'(t_{1})\,,\ \ \ x_{1}=\varphi'(1)\,.
$$
Eqs. (\ref{FiT1}) and (\ref{DerFi01}) lead to
$$
u_{0}=\frac{t_{1}}{z_{1}}\,x_{0},\ \ \ v_{0}= \frac{t_{1}}{z_{1}}\,y_{1},\ \ \ u_{1}=\frac{1-t_{1}}{1-z_{1}}\,y_{1},\ \ \ v_{1}=\frac{1-t_{1}}{1-z_{1}}\,x_{1},
$$
 and
$$
\left|\det \frac{\partial(u_{1},\,v_{1},\,u_{2},\,v_{2})}{\partial (z,\,x_{0},\,x_{1},\,y_{1})} \right|=\frac{\left[t_{1}(1-t_{1})\right]^{2}}{\left[z_{1}(1-z_{1})\right]^{3}}\,y_{1}\,.
$$

Thus, in this case, we get the transparent rule of the functional integration
$$
\int\limits_{Diff^{1}_{+} ([0,\,1]) }\,\Phi\left(\varphi(t_{1});\,\varphi'(0),\,\varphi'(t_{1}),\,\varphi'(1)\right)\ \mu_{\sigma}(d\varphi)
$$
$$
=\left[t_{1}(1-t_{1})\right]^{2}\,\int\limits_{0}^{1}\left[z_{1}(1-z_{1})\right]^{-3}\,dz_{1}\,\int\limits_{0}^{+\infty}\,dx_{0}\,\int\limits_{0}^{+\infty}\,
y_{1}\,dy_{1}\,
\int\limits_{0}^{+\infty}\,dx_{1}
$$
\begin{equation}
   \label{Rule1}
\times\,\Phi\left(z_{1};\,x_{0},\,y_{1},\,x_{1}\right)\,\mathcal {E}_{\sigma\sqrt{t_{1}}}\left(\frac{t_{1}}{z_{1}}\,x_{0},\,\frac{t_{1}}{z_{1}}\,y_{1}\right)\,\mathcal {E}_{\sigma\sqrt{1-t_{1}}}\left(\frac{1-t_{1}}{1-z_{1}}\,y_{1},\,\frac{1-t_{1}}{1-z_{1}}\,x_{1}\right)\,.
\end{equation}

If the integrand $\Phi $ depends on values of the function $\varphi $ and its derivative $\varphi' $ at other points $t_{1},\,t_{2},\,...\,t_{k} $ of the interval, we can continue
 the described procedure
$$
\varphi\rightarrow(\psi_{0},\,\psi_{1})\,,\ \ \psi_{1}\rightarrow(\psi_{2},\,\psi_{3})\,,\ \ \psi_{3}\rightarrow(\psi_{4},\,\psi_{5})\,,\ ...\
\psi_{2k-3}\rightarrow(\psi_{2(k-1)},\,\psi_{2k-1})\,,
$$
and obtain the generalization of (\ref{Rule1})
$$
\int\limits_{Diff^{1}_{+} ([0,\,1]) }\,\Phi\left(\varphi(t_{1}),\,...\,\varphi(t_{k});\,\varphi'(0),\,\varphi'(t_{1}),\,...\,\varphi'(t_{k}),\,\varphi'(1)\right)\ \mu_{\sigma}(d\varphi)
$$
$$
=\left[t_{1}(t_{2}-t_{1})\cdot\cdot\cdot(t_{k}-t_{k-1})(1-t_{k})\right]^{2}\,\int\limits_{0}^{1}dz_{1}...\int\limits_{z_{k-1}}^{1}dz_{k}\,
\left[z_{1}(z_{2}-z_{1})\cdot\cdot\cdot(z_{k}-z_{k-1})(1-z_{k})\right]^{-3}
$$
$$
\times\int\limits_{0}^{+\infty}dx_{0}\int\limits_{0}^{+\infty}
y_{1}\,dy_{1}\cdot\cdot\cdot\int\limits_{0}^{+\infty}
y_{k}\,dy_{k}
\int\limits_{0}^{+\infty}dx_{1}\,\Phi\left(z_{1},\,...\,z_{k};\,x_{0},\,y_{1},\,...\,y_{k},\,x_{1}\right)\,\mathcal {E}_{\sigma\sqrt{t_{1}}}\left(\frac{t_{1}}{z_{1}}\,x_{0},\,\frac{t_{1}}{z_{1}}\,y_{1}\right)
$$
$$
\times\,\mathcal {E}_{\sigma\sqrt{t_{2}-t_{1}}}\left(\frac{t_{2}-t_{1}}{z_{2}-z_{1}}\,y_{1},\,\frac{t_{2}-t_{1}}{z_{2}-z_{1}}\,y_{2}\right)\,\cdot\cdot\cdot\,
\mathcal {E}_{\sigma\sqrt{t_{k}-t_{k-1}}}\left(\frac{t_{k}-t_{k-1}}{z_{k}-z_{k-1}}\,y_{k-1},\,\frac{t_{k}-t_{k-1}}{z_{k}-z_{k-1}}\,y_{k}\right)
$$
\begin{equation}
   \label{RuleK}
\times\,\mathcal {E}_{\sigma\sqrt{1-t_{k}}}\left(\frac{1-t_{k}}{1-z_{k}}\,y_{k},\,\frac{1-t_{k}}{1-z_{k}}\,x_{1}\right)\,.
\end{equation}
In this case, the Jacobian of the substitution is
$$
\left|\det \frac{\partial(u_{1},\,v_{1},\,...\,u_{k+1},\,v_{k+1})}{\partial (z_{1},\,...\,z_{k},\,x_{0},\,y_{1},\,...\,y_{k},\,x_{1})} \right|=\left[t_{1}\cdot\cdot\cdot(1-t_{k})\right]^{2}\,\left[z_{1}\cdot\cdot\cdot(1-z_{k})\right]^{-3}\,y_{1}\cdot\cdot\cdot y_{k}\,.
$$

Note that (\ref{E}) is the only one functional integral we need to evaluate.
In \cite{(BShCorrel)}, we performed the functional integration explicitly and represent
the functional integral (\ref{E}) in the form of the ordinary integral:
$$
\mathcal {E}_{\sigma }\left(u,\,v\right)
=\left(\frac{2}{\pi\sigma^{2}}\right)^{\frac{3}{2}}\,\frac{1}{\sqrt{uv}}\,\exp\left\{\frac{2}{\sigma^{2}}\left(\pi^{2}-u-v\right) \right\}
$$
\begin{equation}
   \label{Eint}
\times\int\limits_{0}^{+\infty}\,\exp\left\{-\frac{2}{\sigma^{2}}\left(2\,\sqrt{uv}\,\cosh\theta+\theta^{2}\right) \right\}\,\sin\left(\frac{4\pi\theta}{\sigma^{2}} \right)\,\sinh(\theta)\,d\theta \,.
\end{equation}

Now the functional integrals (\ref{Rule1}) and (\ref{RuleK}) are reduced to ordinary multiple integrals.

\section { SYK correlation functions}
\label{sec:two}

The renormalized two-point SYK correlation function looks like
$$
<G_{2}^{n}\left(0,\,t_{1} \right)>^{R}_{SYK}
$$
$$
=\frac{2\pi}{\sigma^{2}}\pi^{-\frac{n}{2}}\int\limits_{Diff^{1} ([0,1]) }\delta\left(\frac{\varphi'(1)}{\varphi'(0)}-1 \right)\left(
\frac{\left(\varphi'(t_{1})\varphi'(0)\right)^{\frac{1}{4}}}{\left |\varphi ( t_{1})\right|^{\frac{1}{2}}}\right)^{n}
\exp\left\{\frac{8}{\sigma^{2}}\left(\varphi'(0)\right)\right\}\,\mu_{\sigma}(d\varphi)
$$
$$
= \frac{2\pi}{\sigma^{2}}\pi^{-\frac{n}{2}}\,\left[t_{1}\,(1-t_{1})\right]^{2}\int\limits_{0}^{1}\,dz_{1}\,z_{1}^{-3-\frac{n}{2}}(1-z_{1})^{-3}\, \int\limits_{0}^{+\infty}\int\limits_{0}^{+\infty}dx_{0}dy_{1}\,\left(x_{0}\,y_{1}\right)^{1+\frac{n}{4}}
$$
\begin{equation}
   \label{Ge}
\times\,\exp\left\{\frac{8}{\sigma^{2}}
x_{0} \right\}\mathcal {E}_{\sigma \sqrt{t_{1}}}\left(\frac{t_{1}}{z_{1}}x_{0},\,\frac{t_{1}}{z_{1}}y_{1}\right)\,\mathcal {E}_{\sigma \sqrt{1-t_{1}}}\left( \frac{1-t_{1}}{1-z_{1}} y_{1},\,\frac{1-t_{1}}{1-z_{1}}x_{0}\right)\,.
\end{equation}

In the similar way, we can write the time-ordered (TO) and the out-of-time-ordered (OTO) renormalized four point correlation functions:
$$
<G_{4}^{TO}\left(0,\,t_{1}\,;\ t_{2},\,t_{3} \right)>^{R}_{SYK}
$$
\begin{equation}
   \label{G4TO}
=\frac{2}{\sigma^{2}}\int\limits_{Diff^{1} ([0,1]) }\delta\left(\frac{\varphi'(1)}{\varphi'(0)}-1 \right)\,\frac{\left(\varphi'(t_{1})\varphi'(0)\right)^{\frac{1}{4}}}{\left |\varphi(t_{1})\right|^{\frac{1}{2}}}\,
\frac{\left(\varphi'(t_{3})\varphi'(t_{2})\right)^{\frac{1}{4}}}{\left |\varphi ( t_{3})-\varphi(t_{2})\right|^{\frac{1}{2}}}\,\exp\left\{\frac{8}{\sigma^{2}}\left(\varphi'(0)\right)\right\}\,\mu_{\sigma}(d\varphi)\,,
\end{equation}
and
$$
<G_{4}^{OTO}\left(0,\,t_{2}\,;\ t_{1},\,t_{3} \right)>^{R}_{SYK}
$$
\begin{equation}
   \label{G4OTO}
=\frac{2}{\sigma^{2}}\int\limits_{Diff^{1} ([0,1]) }\delta\left(\frac{\varphi'(1)}{\varphi'(0)}-1 \right)\,\frac{\left(\varphi'(t_{2})\varphi'(0)\right)^{\frac{1}{4}}}{\left |\varphi(t_{2})\right|^{\frac{1}{2}}}\,
\frac{\left(\varphi'(t_{3})\varphi'(t_{1})\right)^{\frac{1}{4}}}{\left |\varphi ( t_{3})-\varphi(t_{1})\right|^{\frac{1}{2}}}\,\exp\left\{\frac{8}{\sigma^{2}}\left(\varphi'(0)\right)\right\}\,\mu_{\sigma}(d\varphi)\,.
\end{equation}
(In the both equations, we assume that $0<t_{1}<t_{2}<t_{3}<1\,.$)

In terms of the functions $\mathcal {E}\,,$ the four-point correlation functions look like
$$
<G_{4}^{TO \,(OTO)}\left(0,\,t_{1\,(2)}\,;\ t_{2\,(1)} ,\,t_{3} \right)>^{R}_{SYK}
=\frac{2}{\sigma^{2}}\,\left[t_{1}(t_{2}-t_{1})(t_{3}-t_{2})(1-t_{3})\right]^{2}\,
$$
$$
\times\,\int\limits_{0}^{1}dz_{1}\int\limits_{z_{1}}^{1}dz_{2}\int\limits_{z_{2}}^{1}dz_{3}\
\chi ^{TO \,(OTO) }\left(z_{1},\,z_{2},\,z_{3}\right)\int\limits_{0}^{+\infty}dx_{0}\int\limits_{0}^{+\infty}
dy_{1}\int\limits_{0}^{+\infty}
dy_{2}\int\limits_{0}^{+\infty}
dy_{3}\,\left[x_{0}\,y_{1}\,y_{2}\,y_{3}\right]^{\frac{5}{4}}
$$
$$
\times\,\exp\left\{\frac{8}{\sigma^{2}}
x_{0} \right\}\,\mathcal {E}_{\sigma\sqrt{t_{1}}}\left(\frac{t_{1}}{z_{1}}\,x_{0},\,\frac{t_{1}}{z_{1}}\,y_{1}\right)\,\mathcal {E}_{\sigma\sqrt{t_{2}-t_{1}}}\left(\frac{t_{2}-t_{1}}{z_{2}-z_{1}}\,y_{1},\,\frac{t_{2}-t_{1}}{z_{2}-z_{1}}\,y_{2}\right)
$$
\begin{equation}
   \label{G4TOOTO}
\times\,
\mathcal {E}_{\sigma\sqrt{t_{3}-t_{2}}}\left(\frac{t_{3}-t_{2}}{z_{3}-z_{2}}\,y_{2},\,\frac{t_{3}-t_{2}}{z_{3}-z_{2}}\,y_{3}\right)\,\mathcal {E}_{\sigma\sqrt{1-t_{3}}}\left(\frac{1-t_{3}}{1-z_{3}}\,y_{3},\,\frac{1-t_{3}}{1-z_{3}}\,x_{0}\right)\,,
\end{equation}
where
\begin{equation}
   \label{chiTO}
\chi ^{TO  }\left(z_{1},\,z_{2},\,z_{3}\right)=
\left[z_{1}\,(z_{2}-z_{1})\,(z_{3}-z_{2})\,(1-z_{3})\right]^{-3}\,
\left[z_{1}\,(z_{3}-z_{2}) \right]^{-\frac{1}{2}}\,,
\end{equation}
and
\begin{equation}
   \label{chiOTO}
\chi ^{OTO }\left(z_{1},\,z_{2},\,z_{3}\right)=
\left[z_{1}\,(z_{2}-z_{1})\,(z_{3}-z_{2})\,(1-z_{3})\right]^{-3}\,
\left[z_{2}\,(z_{3}-z_{1}) \right]^{-\frac{1}{2}}\,.
\end{equation}

Note that the only difference between (\ref{chiTO}) and (\ref{chiOTO}) is in the dependence of the integrands on the variables $z_{i}\,.$

Neither the OTO four-point correlation function nor the TO four-point correlation function has the form of the product of two two-point correlation
functions.

In appendix B, we obtain  the representations for the correlation functions in terms of the ordinary integrals that can be analyzed numerically.

Using the general rule (\ref{RuleK}) we can write any $n-$point correlator in the same way.

\section{Concluding remarks}
\label{sec:concl}

The paper completes the elaboration of the functional integrals calculus in the theories invariant under the groups of diffeomorphisms started in  \cite{(BShExact)} and  \cite{(BShCorrel)}.

The general rules derived here give a straightforward scheme to evaluate functional integrals
in the theories of the Schwarzian type. The great merit of the method is that it reduces  various problems to the evaluation of the functional integral (\ref{E}) only.

The functional integral (\ref{E})
$$
\mathcal {E}_{\sigma}(u,\,v)=\int\limits_{Diff^{1} ([0,1]) }\,\delta\left(\varphi'(0)-u \right)\,\delta\left(\varphi'(1)-v \right)\,\mu_{\sigma}(d\varphi)
$$
is evaluated explicitly. And the result is written in the form of the ordinary integral (\ref{Eint}).
Therefore the evaluations of other functional integrals of the form (\ref{RuleK}) lead also to ordinary (multiple) integrals.

In the introduction, we explain the difference between functional integration over the group $ Diff^{1}_{+}(S^{1})$ and that over the group $Diff^{1}_{+}(\textbf{R})\,.$ Depending on the problem (e.g., closed or open boundary in $AdS_{2}$ is considered), one should decide what kind of functional integrals is adequate.

The functional integrals over $Diff^{1}_{+}(\textbf{R})/P$  corresponding to SYK correlation functions were evaluated in many papers  (see,e.g., \cite{(BAK1)}, \cite{(BAK2)}, \cite{(MTV)}, \cite{(Blomm)}).

As far as we know, the results for correlation functions defined as functional integrals over $Diff^{1}_{+}(S^{1})/SL(2,\textbf{R}) $ has been obtained here for the first time.

The fact that in this case, the four-point correlation functions cannot be represented in the form of the product of two two-point correlation
functions is connected with the apparent non-Markov behaviour of the function $\varphi(t)\,. $
Although $\xi $ is the Wiener process, the Markov property is violated by the nonlocal substitution  (\ref{subst}) .
The non-Markov property of the measure (\ref{MeasureS}) also manifests itself in  (\ref{Ge}) and (\ref{G4TOOTO}).
The regions with $t>t_{1}$ (for $<G_{2}\left(0,\,t_{1} \right)>^{R}_{SYK} $), and with $t>t_{3}$ (for $
<G_{4}\left(0,\,t_{1\,(2)}\,;\ t_{2\,(1)} ,\,t_{3} \right)>^{R}_{SYK}
$ ) give the nonzero inputs into the integrals for correlation functions.

If one considers $t $ as the time variable, then one can say that the present is influenced by the future.
It is a direct consequence of the fact that the ends of the interval are glued together and the theory is $SL(2, \textbf{R})$ invariant.

In this paper, we obtain the correlation functions in the form that can be analyzed numerically.
The analysis is planned to be the subject of another paper.

\appendix

\section{Convolution of the $\mathcal {E} $ functions  }

Let us
take the integrand of (\ref{E}) for $ \Phi$. In this case, we get the convolution rule for two $\mathcal {E} $ functions:
$$
\mathcal {E}_{\sigma}(u,\,v)=\left[t_{1}(1-t_{1})\right]^{2}\,\int\limits_{0}^{1}\left[z_{1}(1-z_{1})\right]^{-3}\,dz_{1}\,\int\limits_{0}^{+\infty}\,
y_{1}\,dy_{1}\,
$$
\begin{equation}
   \label{ConvRule}
\times\,\mathcal {E}_{\sigma\sqrt{t_{1}}}\left(\frac{t_{1}}{z_{1}}\,u,\,\frac{t_{1}}{z_{1}}\,y_{1}\right)\,\mathcal {E}_{\sigma\sqrt{1-t_{1}}}\left(\frac{1-t_{1}}{1-z_{1}}\,y_{1},\,\frac{1-t_{1}}{1-z_{1}}\,v\right)\,.
\end{equation}

We would like to stress that this convolution rule has been proved by direct explicit evaluation of functional integrals.

However, to demonstrate once more
that the dependence of the right-hand side of (\ref{ConvRule}) on $t_{1}$ is fictitious, we consider the integral
\begin{equation}
   \label{IntE}
\int\limits_{0}^{+\infty}\mathcal {E}_{\sigma }(u,\,u)\,u\,du= \frac{1}{\sqrt{2\pi^{3}\sigma^{2}}}\int\limits_{0}^{+\infty}
d\theta\,\exp\left\{-\frac{2}{\sigma^{2}}\left[
\theta^{2}-\pi^{2}\right]\right\}\,
\sin\left(\frac{4\pi\,\theta}{\sigma^{2}}\right)
\tanh(\frac{\theta}{2})\,.
\end{equation}
At the same time, we have from the right-hand side of (\ref{ConvRule})
$$
\frac{2}{\pi^{3}\sigma^{2}}\left[t_{1}\,(1-t_{1})\right]^{-\frac{1}{2}}\int\limits_{0}^{+\infty}\,\int\limits_{0}^{+\infty}d\theta_{1}\, d\theta_{2}\,\exp\left\{-\frac{2}{\sigma^{2}}\left[\frac{\theta_{1}^{2}-\pi^{2}}{t_{1}}+
\frac{\theta_{2}^{2}-\pi^{2}}{(1-t_{1})}\right]\right\}
$$
$$
\times \sin\left(\frac{4\pi\,\theta_{1}}{\sigma^{2}\,t_{1}}\right)
\sin\left(\frac{4\pi\,\theta_{2}}{\sigma^{2}\,(1-t_{1})}\right)\,\sinh \theta_{1}\sinh\theta_{2}\,\int\limits_{0}^{1}\,dz_{1}\,\left[z_{1}(1-z_{1})\right]^{-2}
$$
$$
\times \int\limits_{0}^{+\infty}\int\limits_{0}^{+\infty}\,du\,dy_{1}\,\exp
\left\{-\frac{1}{z_{1}(1-z_{1})}\left(u+y_{1}+2\left[ (1-z_{1})\cosh \theta_{1} + z_{1}\cosh \theta_{2} \right]\sqrt{uy_{1}}\, \right)  \right\}
$$
$$
= \frac{2}{\pi^{3}\sigma^{2}}\left[t_{1}\,(1-t_{1})\right]^{-\frac{1}{2}}\,
\int\limits_{0}^{+\infty}\,\int\limits_{0}^{+\infty}d\theta_{1}\, d\theta_{2}\,\exp\left\{-\frac{2}{\sigma^{2}}\left[\frac{\theta_{1}^{2}-\pi^{2}}{t_{1}}+
\frac{\theta_{2}^{2}-\pi^{2}}{(1-t_{1})}\right]\right\}\sinh\theta_{1}\sinh\theta_{2}$$
\begin{equation}
   \label{ResConv}
\times \,\sin\left(\frac{4\pi\,\theta_{1}}{\sigma^{2}\,t_{1}}\right)
\sin\left(\frac{4\pi\,\theta_{2}}{\sigma^{2}\,(1-t_{1})}\right)\,\int\limits_{0}^{1}\,dz_{1}
\left[\frac{b}{\left(b^{2}-1 \right)^{\frac{3}{2}}}arc\cosh b -\frac{1}{b^{2}-1} \right]\,,
\end{equation}
where
$$
b=\left[(1-z_{1})\cosh\theta_{1} + z_{1} \cosh\theta_{2} \right]\,.
$$
To obtain (\ref{ResConv}), we have re-scaled the variables
$$
\bar{u}=\frac{u}{z_{1}(1-z_{1})}\,,\ \ \bar{y_{1}}=\frac{y_{1}}{z_{1}(1-z_{1})}\,.
$$
And then after the substitutions
\begin{equation}
   \label{substitutions}
\bar{u}=\rho\cos^{2}\omega\,,\ \ \bar{y_{1}}=\rho\sin^{2}\omega\,,\ \ \ \ \xi=\tan \omega \,,
\end{equation}
we have used the table of integrals \cite{(Prud)} (eq. 2.2.9.15).

The results of computations of (\ref{ResConv})  at ten different points of the interval $[0,\,1]$ coincide  with the result of computation of (\ref{IntE})
 with very high accuracy
 (e.g., the results are 0.07978 for $\sigma=5$, and 0.03989 for $\sigma=10$ ).

\section{Correlation functions in terms of ordinary integrals  }

In terms of the ordinary integrals, the two-point correlation function (\ref{Ge}) has the form
$$
<G_{2}^{n}\left(0,\,t_{1} \right)>^{R}_{SYK}=\frac{1}{\pi^{2}}\left(\frac{\sigma^{2}}{2\pi}\right)^{\frac{n}{2}}\,
\left[t_{1}(1-t_{1})\right]^{-\frac{1}{2}}
$$
$$
\times\int\limits_{0}^{+\infty}\,\int\limits_{0}^{+\infty}d\theta_{1}\, d\theta_{2}\,\exp\left\{-\frac{2}{\sigma^{2}}\left[\frac{\theta_{1}^{2}-\pi^{2}}{t_{1}}+
\frac{\theta_{2}^{2}-\pi^{2}}{(1-t_{1})}\right]\right\}
\,
\sin\left(\frac{4\pi\,\theta_{1}}{\sigma^{2}\,t_{1}}\right)
\sin\left(\frac{4\pi\,\theta_{2}}{\sigma^{2}\,(1-t_{1})}\right)
$$
$$
\times \sinh(\theta_{1})\sinh(\theta_{2})\,\int\limits_{0}^{1}\,dz_{1}\,(1-z_{1})^{\frac{n}{2}}\,
\int\limits_{0}^{+\infty}\int\limits_{0}^{+\infty}\,dx_{0}\,dy_{1}\,\left(x_{0}\,y_{1}\right)^{\frac{n}{4}}
$$
\begin{equation}
   \label{J1}
\times \exp
\left\{-\left(x_{0}+y_{1}+2\left[ (1-z_{1})\cosh \theta_{1} + z_{1}\cosh \theta_{2} \right]\sqrt{x_{0}y_{1}}\, -4z_{1}(1-z_{1})x_{0} \right)\right\}\,.
\end{equation}

After the substitutions (\ref{substitutions}), the integrals over $x_{0}$ and $y_{1}$ are transformed into the integral
\begin{equation}
   \label{Intksi}
\int\limits_{0}^{+\infty}\frac{2\,\xi^{1+\frac{n}{2}}}{\left[\xi^{2}+2b\xi+1-4z_{1}(1-z_{1}) \right]^{2+\frac{n}{2}}}\,d\xi\,.
\end{equation}
It is convenient to use the table of integrals \cite{(Prud)} (eq. 2.2.9.8).
$$
<G_{2}^{n}\left(0,\,t_{1} \right)>^{R}_{SYK}=\frac{2}{\pi^{2}}\left(\frac{\sigma^{2}}{2\pi}\right)^{\frac{n}{2}}\,\Gamma^{3}\left(\frac{n+4}{2}\right)\,\Gamma^{-1}\left(n+4\right)\,
\left[t_{1}(1-t_{1})\right]^{-\frac{1}{2}}
$$
$$
\times\int\limits_{0}^{+\infty}\,\int\limits_{0}^{+\infty}d\theta_{1}\, d\theta_{2}\,\exp\left\{-\frac{2}{\sigma^{2}}\left[\frac{\theta_{1}^{2}-\pi^{2}}{t_{1}}+
\frac{\theta_{2}^{2}-\pi^{2}}{(1-t_{1})}\right]\right\}
\,
\sin\left(\frac{4\pi\,\theta_{1}}{\sigma^{2}\,t_{1}}\right)
\sin\left(\frac{4\pi\,\theta_{2}}{\sigma^{2}\,(1-t_{1})}\right)
$$
$$
\times \sinh(\theta_{1})\sinh(\theta_{2})\,\int\limits_{0}^{1}\,dz_{1}\,(1-z_{1})^{\frac{n}{2}}\,
\left[1- 4 z_{1}(1-z_{1})\right]^{-\frac{n+4}{4}}
$$
\begin{equation}
   \label{nodd}
\times \,F\left(\frac{n+4}{2},\,\frac{n+4}{2};\,\frac{n+5}{2};\,\frac{1}{2}\left(1-\frac{b}{1- 4 z_{1}(1-z_{1})}\right)\, \right)\,,
\end{equation}
where $F$ is the Gaussian hypergeometric function, and
\begin{equation}
   \label{b}
b=(1-z_{1})\cosh \theta_{1}+ z_{1}\cosh \theta_{2}\,.
\end{equation}

For $n=2k,\ \ \ k=0,\,1\,...\,,$ we can also use  eq. 2.2.9.12 of \cite{(Prud)}, and rewrite the result in the form
$$
<G_{2}^{n}\left(0,\,t_{1} \right)>^{R}_{SYK}=\left(-1\right)^{k+1}\frac{1}{(k+1)!}\frac{1}{2\pi^{2}}\left(\frac{\sigma^{2}}{4\pi}\right)^{k}\Gamma\left(k+2\right)\,
\left[t_{1}(1-t_{1})\right]^{-\frac{1}{2}}
$$
$$
\times\int\limits_{0}^{+\infty}\,\int\limits_{0}^{+\infty}d\theta_{1}\, d\theta_{2}\,\exp\left\{-\frac{2}{\sigma^{2}}\left[\frac{\theta_{1}^{2}-\pi^{2}}{t_{1}}+
\frac{\theta_{2}^{2}-\pi^{2}}{(1-t_{1})}\right]\right\}
\,
\sin\left(\frac{4\pi\,\theta_{1}}{\sigma^{2}\,t_{1}}\right)
\sin\left(\frac{4\pi\,\theta_{2}}{\sigma^{2}\,(1-t_{1})}\right)
$$
$$
\times \sinh(\theta_{1})\sinh(\theta_{2})\,\int\limits_{0}^{1}\,dz_{1}\,(1-z_{1})^{k}\,
$$
\begin{equation}
   \label{even}
\times \,\frac{\partial^{k+1}}{\partial b^{k+1}}\left(\frac{1}{\sqrt{b^{2}-1+ 4 z_{1}(1-z_{1})}}\log\,\frac{b+\sqrt{b^{2}-1+ 4 z_{1}(1-z_{1})}}{b-\sqrt{b^{2}-1+ 4 z_{1}(1-z_{1})}}\right)\,,
\end{equation}
where $b$ is given by (\ref{b}).

Here, we also present the four-point correlation functions in the form that can be analyzed numerically. It is
$$
<G_{4}^{TO \,(OTO)}\left(0,\,t_{1\,(2)}\,;\ t_{2\,(1)} ,\,t_{3} \right)>^{R}_{SYK}
=\left(\frac{2}{\pi^{3}\sigma^{2}}\right)^{2}\,\left[t_{1}(t_{2}-t_{1})(t_{3}-t_{2})(1-t_{3})\right]^{-\frac{1}{2}}
$$
$$
\times\,\int\limits_{0}^{1}dz_{1}\int\limits_{z_{1}}^{1}dz_{2}\int\limits_{z_{2}}^{1}dz_{3}\, \
\tilde{\chi} ^{TO \,(OTO) }\left(z_{1},\,z_{2},\,z_{3}\right)
$$
$$
\times\,\int\limits_{0}^{+\infty}\,d\theta_{1}...\int\limits_{0}^{+\infty}\, \,d\theta_{4}\,\exp\left\{-\frac{2}{\sigma^{2}}\left[\frac{\theta_{1}^{2}-\pi^{2}}{\tau_{1}}+
\frac{\theta_{2}^{2}-\pi^{2}}{(\tau_{2}-\tau_{1})}+
\frac{\theta_{3}^{2}-\pi^{2}}{(\tau_{3}-\tau_{2})}+
\frac{\theta_{4}^{2}-\pi^{2}}{(1-\tau_{3})}\right]\right\}
$$
$$
\times\,
\sin\left(\frac{4\pi\,\theta_{1}}{\sigma^{2}\,\tau_{1}}\right)\,
\sin\left(\frac{4\pi\,\theta_{2}}{\sigma^{2}\,(\tau_{2}-\tau_{1})}\right)\,\sin\left(\frac{4\pi\,\theta_{3}}{\sigma^{2}\,(\tau_{3}-\tau_{2})}\right)\,
$$
$$
\times\,\sinh\theta_{1}\,\sinh\theta_{2}\,\sinh\theta_{3}\,\sinh\theta_{4}
\int\limits_{0}^{+\infty}dx_{0}\int\limits_{0}^{+\infty}
dy_{1}\int\limits_{0}^{+\infty}
dy_{2}\int\limits_{0}^{+\infty}
dy_{3}\,\left[x_{0}\,y_{1}\,y_{2}\,y_{3}\right]^{\frac{1}{4}}
$$
$$
\times \exp\left\{4x_{0}-\frac{1}{z_{1}}\left[x_{0}+y_{1}+2\sqrt{x_{0}y_{1}}\cosh \theta_{1} \right]
-\frac{1}{z_{2}-z_{1}}\left[y_{1}+y_{2}+2\sqrt{y_{1}y_{2}}\cosh \theta_{2} \right]\right\}
$$
\begin{equation}
\label{4int}
\times \exp\left\{-\frac{1}{z_{3}-z_{2}}\left[y_{2}+y_{3}+2\sqrt{y_{2}y_{3}}\cosh \theta_{3} \right]
-\frac{1}{1-z_{3}}\left[y_{3}+x_{0}+2\sqrt{y_{3}x_{0}}\cosh \theta_{4} \right] \right\}\,,
\end{equation}
where
$$
\tilde{\chi} ^{TO  }\left(z_{1},\,z_{2},\,z_{3}\right)=
\left[z_{1}\,(z_{2}-z_{1})\,(z_{3}-z_{2})\,(1-z_{3})\right]^{-2}\,
\left[z_{1}\,(z_{3}-z_{2}) \right]^{-\frac{1}{2}}\,,
$$
and
$$
\tilde{\chi} ^{OTO }\left(z_{1},\,z_{2},\,z_{3}\right)=
\left[z_{1}\,(z_{2}-z_{1})\,(z_{3}-z_{2})\,(1-z_{3})\right]^{-2}\,
\left[z_{2}\,(z_{3}-z_{1}) \right]^{-\frac{1}{2}}\,.
$$

\section{Functional integrals over the space $Diff^{1}_{+}(\textbf{R}) $ }

In this appendix, we evaluate functional integrals over $Diff^{1}_{+}(\textbf{R}) $ of the form
\begin{equation}
   \label{NewFI}
\int \limits_{Diff^{1}_{+}(\textbf{R}) }\,\Psi \left(f(\tau_{1}),\,f(\tau_{2}),\,... \right)\
 \exp\left\{\frac{1}{\sigma^{2}}\int \limits _{-\infty}^{+\infty}\,\mathcal{S}_{f}(\tau)\,d\tau\, \right\}\,df\,.
\end{equation}

Consider the two-point correlation function in this theory
$$
\mathcal {G}_{2}^{n}(0,\,\tau)=\int \limits_{Diff^{1}_{+}(\textbf{R})/P}\,\frac{\left(f'(0)f'(\tau)\right)^{\frac{n}{4}}}{\left |f( \tau)-f(0)\right|^{\frac{n}{2}}}\
 \exp\left\{\frac{1}{\sigma^{2}}\int \limits _{-\infty}^{+\infty}\,\mathcal{S}_{f}(t)\,dt\, \right\}\,df\,.
$$

As it is explained in the introduction, we should factorize the integration space. To exclude the non-relevant degrees of freedom in another way, we can fix
the values
$$
f(0)=0\,, \ \ f'(0)=1\,.
$$

After the substitution $f(\tau)=\int\limits^{\tau}_{0}\exp\{\xi(t)\}dt\,,\ \ \xi(0)=0$ (\cite{(BAK1)}, \cite{(BAK2)}),  it looks like
$$
\mathcal {G}_{2}^{n}(0,\,\tau)=\int \limits_{C((-\infty,\,+\infty))}\,\frac{e^{\frac{n}{4}(\xi(\tau))}}{\left(\int \limits_{0}^{\tau}\,e^{\xi(t)}dt\right)^{\frac{n}{2}}}
 \exp\left\{-\frac{1}{2\sigma^{2}}\int \limits _{-\infty}^{+\infty}\,\left(\xi'(t)\right)^{2}\,dt\, \right\}\,d\xi\,.
$$
Due to the Markov property of the Wiener measure, it can be rewritten as
\begin{equation}
   \label{Bksi}
\mathcal {G}_{2}^{n}(0,\,\tau)=\int \limits_{C([0,\,\tau])}\,\frac{e^{\frac{n}{4}(\xi(\tau))}}{\left(\int \limits_{0}^{\tau}\,e^{\xi(t)}dt\right)^{\frac{n}{2}}}
 \exp\left\{-\frac{1}{2\sigma^{2}}\int \limits_{0}^{\tau}\,\left(\xi'(t)\right)^{2}\,dt\, \right\}\,d\xi\,.
\end{equation}

Instead of transforming the Wiener integral (\ref{Bksi}) to the functional integral in Liouville theory as it was proposed in \cite{(BAK1)}, \cite{(BAK2)}, we use the method described above.
First, we re-scale the time variable $t=\tau\,\bar{t}$ and substitute $\eta(\bar{t})=\xi(t)\,. $
Then (\ref{Bksi}) has the form
\begin{equation}
   \label{Beta}
\mathcal {G}_{2}^{n}(0,\,\tau)=\tau^{-\frac{n}{2}}\,\int \limits_{C([0,\,1])}\,\frac{e^{\frac{n}{4}\eta(1)}}{\left(\int \limits_{0}^{1}\,e^{\eta(\bar{t})}d\bar{t}\right)^{\frac{n}{2}}}
 \exp\left\{-\frac{1}{2\sigma^{2}\tau}\int \limits_{0}^{1}\,\left(\eta'(\bar{t})\right)^{2}\,d\bar{t}\, \right\}\,d\eta\,.
\end{equation}

In terms of the function
$$
\varphi(t)=\frac{\int \limits _{0}^{t}\,e^{\eta(\bar{t})}d\bar{t}}{\int \limits _{0}^{1}\,e^{\eta(\bar{t})}d\bar{t} }\,,
$$
(\ref{Beta}) is written as
\begin{equation}
   \label{BintDiff}
\tau^{-\frac{n}{2}}\,\int \limits_{Diff^{1}_{+}([0,\,1])}\,\left(\varphi'(0) \varphi'(1) \right)^{\frac{n}{4}}\,\mu _{\sigma\sqrt{\tau}}(d\varphi)
=\tau^{-\frac{n}{2}}\,\int \limits_{0}^{+\infty}\,\int \limits_{0}^{+\infty}\,\left(uv \right)^{\frac{n}{4}}\,\mathcal {E} _{\sigma\sqrt{\tau}}(u\,,v)\,du\,dv\,.
\end{equation}

Integrating over $u$ and $v$ as in the appendix A, we obtain
$$
\mathcal {G}_{2}^{n}(0,\,\tau)=2\pi^{-\frac{3}{2}}\left( \frac{\sigma^{2}}{2}\right)^{\frac{n-1}{2}}\,B\left(\frac{n+1}{2},\, \frac{n+1}{2}\right)\,\tau^{-\frac{1}{2}}\,\int\limits_{0}^{+\infty}\,
d\theta\,\sin\left(\frac{4\pi\,\theta}{\sigma^{2}\tau}\right)
\sinh \theta
$$
\begin{equation}
\label{gnResult}
\times \exp\left\{-\frac{2}{\sigma^{2}\tau}\left[
\theta^{2}-\pi^{2}\right]\right\}\,
F\left(\frac{n+1}{2},\, \frac{n+1}{2};\,\frac{n+1}{2}+1;\,\frac{1-\cosh \theta}{2}\right)\,.
\end{equation}

In the case $n=2$,  it looks like
\begin{equation}
\label{g2Result}
\mathcal {G}_{2}^{2}(0,\,\tau)=\left( \frac{\sigma^{2}}{2\pi^{3}}\right)^{\frac{1}{2}}\tau^{-\frac{1}{2}}\int\limits_{0}^{+\infty}
d\theta\exp\left\{-\frac{2}{\sigma^{2}\tau}\left[
\theta^{2}-\pi^{2}\right]\right\}
\sin\left(\frac{4\pi\theta}{\sigma^{2}\tau}\right)
\left[\frac{\theta \cosh\theta }{\sinh^{2}\theta }-\frac{1}{\sinh\theta }\right].
\end{equation}

The asymptotic form of $\mathcal {G}_{2}^{n}(0,\,\tau)$ at $\tau\rightarrow\infty\,$ ,
$$
\left(\mathcal {G}_{2}^{n}(0,\,\tau)\right)_{As}= 4\pi^{-\frac{1}{2}}\left( \frac{\sigma^{2}}{2}\right)^{\frac{n-3}{2}}\,B\left(\frac{n+1}{2},\, \frac{n+1}{2}\right)\ \tau^{-\frac{3}{2}}
$$
\begin{equation}
\label{gnAs}
\times\,\int\limits_{0}^{+\infty}
\,
\theta\,
\sinh \theta\, F\left(\frac{n+1}{2},\, \frac{n+1}{2};\,\frac{n+1}{2}+1;\,\frac{1-\cosh \theta}{2}\right)\,d\theta\,,
\end{equation}
immediately follows from (\ref{gnResult}).

In the same way, we can explicitly evaluate functional integrals for other correlation functions in this theory.

In particular, for
$\mathcal {G}_{4}^{TO}(0,\,\tau_{_{1}};\,\tau_{2},\,\tau_{3})$
 we have
$$
\mathcal {G}_{4}^{TO}(0,\,\tau_{_{1}};\,\tau_{2},\,\tau_{3})=\int \limits_{Diff^{1}_{+}(\textbf{R})/P }\,\frac{\left(f'(0)f'(\tau_{1})\right)^{\frac{1}{4}}}{\left |f( \tau_{1})-f(0)\right|^{\frac{1}{2}}}\,\frac{\left(f'(\tau_{2})f'(\tau_{3})\right)^{\frac{1}{4}}}{\left |f( \tau_{3})-f(\tau_{2})\right|^{\frac{1}{2}}}
 \exp\left\{\frac{1}{\sigma^{2}}\int \limits _{-\infty}^{+\infty}\,\mathcal{S}_{f}(t)\,dt\, \right\}\,df
$$
\begin{equation}
\label{gTO}
=\mathcal {G}_{2}(0,\,\tau_{_{1}})\,\mathcal {G}_{2}(\,\tau_{2},\,\tau_{3})\,.
\end{equation}

The OTO four-point correlation function is given by $(0<\tau_{1}<\tau_{2}<\tau_{3})$
$$
\mathcal {G}_{4}^{OTO}(0,\,\tau_{_{2}};\,\tau_{1},\,\tau_{3})=\int \limits_{Diff^{1}_{+}(\textbf{R})/P }\,\frac{\left(f'(0)f'(\tau_{2})\right)^{\frac{1}{4}}}{\left |f( \tau_{2})-f(0)\right|^{\frac{1}{2}}}\,\frac{\left(f'(\tau_{1})f'(\tau_{3})\right)^{\frac{1}{4}}}{\left |f( \tau_{3})-f(\tau_{1})\right|^{\frac{1}{2}}}
 \exp\left\{\frac{1}{\sigma^{2}}\int \limits _{-\infty}^{+\infty}\,\mathcal{S}_{f}(t)\,dt\, \right\}\,df
$$
$$
=\pi^{\frac{9}{2}}\left(\frac{\sigma^{2}}{2}\right)^{\frac{1}{2}}\left[\tau_{1}(\tau_{2}-\tau_{1})(\tau_{3}-\tau_{2})\right]^{-\frac{1}{2}}
\int\limits_{0}^{1}dz_{1}\int\limits_{z_{1}}^{1}dz_{2}\,
\left[z_{1}(z_{2}-z_{1})(1-z_{2})\right]^{-2}\left[z_{2}(1-z_{1}) \right]^{-\frac{1}{2}}
$$
$$
\times\int\limits_{0}^{+\infty}\,\int\limits_{0}^{+\infty}\,\int\limits_{0}^{+\infty}d\theta_{1}\, d\theta_{2}\,d\theta_{3}\,\exp\left\{-\frac{2}{\sigma^{2}}\left[\frac{\theta_{1}^{2}-\pi^{2}}{\tau_{1}}+
\frac{\theta_{2}^{2}-\pi^{2}}{(\tau_{2}-\tau_{1})}+
\frac{\theta_{3}^{2}-\pi^{2}}{(\tau_{3}-\tau_{2})}\right]\right\}
$$
$$
\times\,
\sin\left(\frac{4\pi\,\theta_{1}}{\sigma^{2}\,\tau_{1}}\right)\,
\sin\left(\frac{4\pi\,\theta_{2}}{\sigma^{2}\,(\tau_{2}-\tau_{1})}\right)\,\sin\left(\frac{4\pi\,\theta_{3}}{\sigma^{2}\,(\tau_{3}-\tau_{2})}\right)\, \sinh\theta_{1}\,\sinh\theta_{2}\,\sinh\theta_{3}
$$
$$
\times\int\limits_{0}^{+\infty}dx_{0}\int\limits_{0}^{+\infty}
dy_{1}\int\limits_{0}^{+\infty}
dy_{2}\int\limits_{0}^{+\infty}
dx_{1}\left[x_{0}x_{1}\right]^{-\frac{1}{4}}  \left[y_{1}y_{2}\right]^{\frac{1}{4}}\exp\left\{
-\frac{1}{z_{1}}\left[x_{0}+y_{1}+2\sqrt{x_{0}y_{1}}\cosh \theta_{1} \right]\right\}
$$
\begin{equation}
\label{gOTO}
\times \exp\left\{
-\frac{1}{z_{2}-z_{1}}\left[y_{1}+y_{2}+2\sqrt{y_{1}y_{2}}\cosh \theta_{2} \right]
-\frac{1}{1-z_{2}}\left[y_{2}+x_{1}+2\sqrt{y_{2}x_{1}}\cosh \theta_{3} \right] \right\}\,.
\end{equation}

In contrast to the correlation functions in the theory considered in the main part of the paper, the correlation functions (\ref{Bksi}), (\ref{gTO}), (\ref{gOTO}) have no input from the regions $t>\tau\,,$ or $t>\tau_{3}\,.$

\end{document}